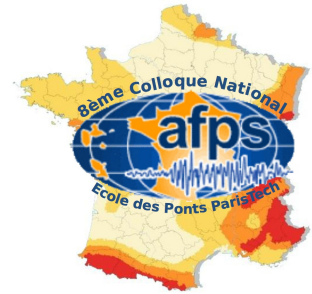

# Evaluation des différents paramètres utilisés pour l'estimation du contenu en fréquences des mouvements du sol, avec application aux forts tremblements de terre de Vrancea, Roumanie

**Iolanda-Gabriela Craifaleanu\*,\*\***

*\* Université Technique de Constructions de Bucarest,*

*Département de constructions en béton armé*

*122-124 Blvd. Lacul Tei, RO-020396 Bucarest, Roumanie*

*i.craifaleanu@gmail.com*

*\*\* Institut National de Recherche et Développement dans la Construction, l'Urbanisme et le*

*Développement Territorial Soutenable – « URBAN – INCERC », Branche INCERC Bucarest, Roumanie*

*266 Sos. Pantelimon, RO-021652 Bucarest, Roumanie*

RÉSUMÉ. *L'article présente des résultats d'une étude détaillée des mouvements sismiques enregistrés pendant les forts tremblements de terre de Vrancea, Roumanie (magnitude de moment $M_w > 6$), qui se sont produits dans les dernières 34 années. En analysant plus de 300 accélérogrammes, on compare la capacité des différentes expressions utilisées dans la littérature d'estimer la période prédominante d'un mouvement sismique, en décelant également les corrélations entre les valeurs obtenues par des différentes évaluations. On analyse la dépendance de la période prédominante du mouvement sismique des divers facteurs d'influence. On compare les paramètres calculés pour les mouvements enregistrés au même séisme sur des différents emplacements, ainsi que pour des mouvements enregistrés sur le même site aux séismes successifs. Les résultats sont interprétés en corrélation avec les informations fournies par les paramètres de la largeur de bande de fréquences. On fait des considérations sur la mesure dans laquelle on peut séparer l'influence, sur le contenu de fréquences, de la source et des conditions géologiques locales, pour des mouvements sismiques enregistrés sur des différents emplacements.*

ABSTRACT. *The paper presents results of a comprehensive study of ground motions recorded during the strong earthquakes (moment magnitude $M_w > 6$) generated during last 34 years by the seismic source of Vrancea, Romania. By analysing over 300 accelerograms, the capacity of different expressions in the literature to estimate the predominant period of a ground motion is compared. The correlation between the values obtained from different evaluations is assessed as well. The dependence of the predominant period of different factors of influence is analysed. Comparisons are made between the parameters determined for the same seismic event at different stations, as well as for ground motions recorded on the same site at successive earthquakes. The results are interpreted in correlation with the information provided by frequency bandwidth parameters. Considerations are made on the measure in which the influence on the frequency content of the source and of local geological conditions can be separated, for seismic motions recorded on different locations.*

MOTS-CLÉS : *contenu en fréquences, période prédominante, mouvements du sol, forts tremblements de terre, indicateurs de la largeur de bande de fréquences.*

## 1. Introduction

L'évaluation de la période prédominante du mouvement sismique présente un intérêt particulier pour les séismes de Vrancea, compte tenu des aspects spécifiques signalés par les recherches menées jusqu'au présent. On a constaté, suite à ces recherches, que pour une très grande partie du territoire affecté par les séismes étudiés il est difficile d'identifier la profondeur du substratum rocheux, en raison des particularités stratigraphiques. Cette difficulté apparait due à l'absence d'un interface où on peut identifier un contraste évident dans la vitesse de propagation des ondes S, dans le sens de l'augmentation signifiante de la vitesse avec la profondeur (Sandi et al., 2004a).

Des études paramétriques parallèles menées, dans la référence citée, sur la fonction de transfert du paquet stratigraphique superficiel de deux emplacements caractéristiques (les stations sismiques Mairie Cernavoda et INCERC Bucarest), ont révélé qu'on peut classifier les sites conformément aux deux situations distinctes :

- la présence d'une forte augmentation de la vitesse de propagation des ondes S à profondeurs faibles ;
- une augmentation graduelle de la vitesse de propagation des ondes S, jusqu'aux profondeurs importantes, à mesure qu'on prend plusieurs couches stratigraphiques dans le calcul.

Dans le premier cas, l'effet de site a une influence signifiante et on trouve une tendance de stabilité des périodes prédominantes des mouvements sismiques.

Dans le second cas, on a constaté une tendance de variabilité du contenu en fréquences du mouvement sismique, en concluant que les influences déterminantes sont, le plus probablement, celles du mécanisme de foyer et des caractéristiques de la radiation / atténuation sur des longues distances.

Le site de l'INCERC est intéressant en ce qu'il appartient à la seconde catégorie et que sa situation est caractéristique pour une très grande partie des zones affectées par les séismes de Vrancea. Ce site est important aussi parce qu'on y a enregistré la plus étroite bande de fréquences et la plus longue période prédominante des mouvements sismiques, en réponse aux forts séismes de Vrancea de 4 mars 1977 et de 30 août 1986.

Une autre conclusion importante révélée, dans l'étude citée, pour le site d'INCERC Bucarest, est que, en considérant des couches stratigraphiques jusqu'à une profondeur de 2800 m, les pics de la fonction de transfert montrent une correspondance remarquable avec les résultats de l'analyse spectrale des valeurs instrumentales. Ainsi, les pics de la fonction de transfert apparaissent aux périodes très proches de celles des pics principaux des spectres de réponse calculés pour les mouvements sismiques enregistrées, sur le site considéré, aux séismes des 4 mars 1977, 30 août 1986 et 30 mai 1990. On remarque aussi, pour le même site, une tendance de stabilisation de la forme de la fonction de transfert une fois qu'on dépasse l'interface ayant un contraste plus accentué des vitesses de propagation des ondes S, située a 600 m de profondeur.

Une recherche basée sur une approche différente (Lungu et. al, 1999) a obtenu, pour le même emplacement d'INCERC Bucarest, des résultats compatibles, en partie, avec les conclusions citées ci-dessus.

On a mesuré, dans un forage d'environ 128 m de profondeur, les vitesses de propagation des ondes S et les caractéristiques des couches stratigraphiques. En utilisant ces données, on a déterminé la période fondamentale du site et on l'a comparé avec les périodes prédominantes obtenues à l'aide de la densité spectrale de puissance et de la fonction d'autocorrélation, pour les enregistrements des séismes du 4 mars 1977 et du 30 août 1986. Les périodes prédominantes sont, pour ces deux séismes, de 1,6 s et, respectivement, de 1,4 s. On n'atteint des valeurs comparables qu'en considérant, pour la période fondamentale du site, l'entier profile stratigraphique. On en déduit que la prise en compte d'un profil stratigraphique d'une profondeur insuffisante conduit à une sous - estimation de la période fondamentale du site.

Outre les problèmes issus de la difficulté d'identifier la position du substratum rocheux pour une partie importante du territoire affecté par la source sismique de Vrancea, l'estimation de l'influence des divers facteurs sur le contenu en fréquences des mouvements sismiques est rendue encore plus difficile à cause des différences très prononcées entre les caractéristiques (mécanisme de foyer, radiation / atténuation) des forts séismes enregistrés après 4 mars 1977, à savoir ceux du 30 août 1986, du 30 mai 1990 et du 31 mai 1990 (Sandi et al., 2004b, Craifaleanu et al., 2006 etc.).

Tenant compte de ces aspects, on a choisi, dans l'étude présentée ici, de réaliser une évaluation comparative des estimations du contenu en fréquences des mouvements sismiques enregistrées aux forts séismes de Vrancea, basée sur plusieurs paramètres de la littérature.

## 2. Evaluation des paramètres utilisés pour décrire le contenu en fréquences des mouvements sismiques

### 2.1. *Prémisses générales*

La période prédominante est un paramètre caractéristique du contenu en fréquences d'un mouvement sismique. On analyse, dans cette section, les corrélations entre les valeurs de la période prédominante, calculées par des différentes expressions proposées dans la littérature, en tenant compte, en même temps, de l'information fournie par les paramètres de la largeur de la bande de fréquences.

Dans l'étude présentée, on à utilisé plus de 300 accélérogrammes enregistrées lors des forts séismes de Vrancea (magnitude de moment $M_w > 6$) qui se sont produits en Roumanie à partir de 1977, à savoir les séismes du 4 mars 1977 (magnitude de moment $M_w = 7,4$, profondeur hypocentrale $h = 109$ km), du 30 août 1986 ($M_w = 7,1$, $h = 133$ km), du 30 mai 1990 ($M_w = 6,9$, $h = 91$ km) et du 31 mai 1990 ($M_w = 6,4$, $h = 87$ km) (INFP et INCERC, 2001).

Pour ces enregistrements sismiques, on a déterminé 3 catégories distinctes de paramètres, à savoir : (a) paramètres basées sur la densité spectrale de puissance et sur le spectre d'amplitude de Fourier, (b) paramètres basées sur les spectres de réponse, et (c) paramètres basées sur les valeurs maximales du mouvement du sol. Une analyse détaillée des corrélations entre ces paramètres a été réalisé dans (Craifaleanu, 2011) ; dans la suite on va présenter une sélection des résultats le plus signifiants.

La distribution des enregistrements considérés dans l'étude, par rapport à l'événement sismique et au type de composante (horizontale ou verticale) est présentée dans la figure 1. A remarquer qu'un seul accélérogramme complet a été enregistré lors du séisme du 4 mars 1977.

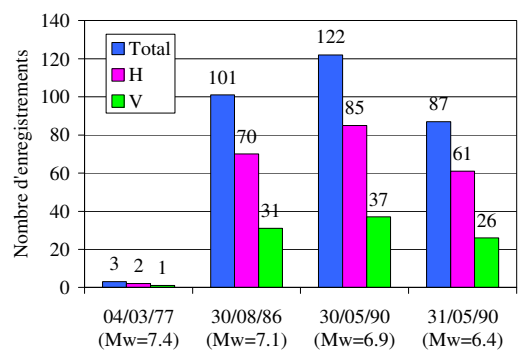

**Figure 1.** *Distribution des enregistrements sismiques considérés dans l'étude, par rapport à l'événement sismique et au type de composante (H − horizontale, V − verticale)*

## 2.2. *Paramètres analysés*

*2.2.1. Paramètres basés sur la densité spectrale de puissance et sur le spectre d'amplitude de Fourier*

(a) Les fréquences $f_1$, $f_2$ et $f_3$, correspondant aux trois premières pics, par ordre d'amplitude, de la densité spectrale de puissance, DSP (JCSS, 2001) ; alternativement, on peut utiliser les périodes $T_1$, $T_2$ et $T_3$. Dans la présente étude, on utilise, pour $T_1$, la notation $T_1^{(DSP)}$.

(b) Les fréquences fractile $f_{10}$, $f_{50}$ et $f_{90}$ correspondant à 10%, 50% et 90% de la DSP cumulative.

(c) La fréquence centrale, $\Omega$ (Vanmarcke, 1976), exprimée par

$$\Omega = \sqrt{\lambda_2/\lambda_0} \qquad [1]$$

La fréquence centrale est une mesure de la fréquence où est concentrée la DSP. Dans la formule [1], $\lambda_0$ et $\lambda_2$ sont les moments spectraux d'ordres 0 et 2 de la DSP unilatérale du processus aléatoire de l'accélération du sol (JCSS, 2001), donnée par la formule générale :

$$\lambda_n = \int_0^{\omega_n} \omega^n G(\omega) d\omega \qquad [2]$$

où $n = 0$ et 2; $\omega$ est la pulsation et $G(\omega)$ est la fonction de densité spectrale. Sur la base de cette définition, on a calculé, pour la présente étude, une période centrale

$$T_{cen} = 2\pi/\Omega \qquad [3]$$

(d) La fréquence moyenne (Thrainsson, 2000), exprimée par

$$\overline{\omega} = \lambda_1/\lambda_0 \qquad [4]$$

Sur la base de cette définition, on a calculé, pour la présente étude, une période moyenne

$$T_{mean} = 2\pi/\overline{\omega} \qquad [5]$$

(e) Le facteur de forme, *q* (Vanmarcke, 1976), est défini par la formule

$$q = \sqrt{1 - \frac{\lambda_1^2}{\lambda_0 \lambda_2}}, \quad 0 \le q \le 1 \qquad [6]$$

Le facteur de forme est utilisé, en général, comme une mesure de la largeur de bande de fréquences, car il reflète la dispersion de la fonction de la densité spectrale de puissance par rapport à la fréquence centrale. Les valeurs plus hautes de ce facteur correspondent aux largeurs de bande plus élevées.

(f) Le paramètre ε de Cartwright et Longuet-Higgins (Cartwright et al., 1956), donné par l'expression [7], est aussi utilisé comme mesure de la largeur de bande de fréquences (JCSS, 2001) :

$$\varepsilon = \sqrt{1 - \frac{\lambda_2^2}{\lambda_0 \lambda_4}}, \quad 0 \le \varepsilon \le 1 \qquad [7]$$

On considère que, pour les processus à large bande de fréquences, les valeurs de ε sont proches de 2/3 et inférieures à 0,85, tandis que pour les processus à bande étroite de fréquences les valeurs de ε sont supérieures à 0,95 (JCSS, 2001).

(g) La période moyenne quadratique, $T_{ms}$, proposée par Rathje et al. (Rathje et al., 1998), est définie comme

$$T_{ms} = \left[ \sum_i C_i^2 \left( 1/f_i \right) \Big/ \sum_i C_i^2 \right] \text{ pour } 0,25\text{Hz} \leq f_i \leq 20\text{Hz} \quad [8]$$

(h) où $C_i$ sont les amplitudes Fourier de l'accélérogramme et $f_i$ sont les fréquences de la transformation rapide de Fourier (FFT), dans le domaine 0,25…20 Hz. Pour obtenir une valeur stable de $T_{ms}$, l'intervalle de fréquences utilisé dans le calcul de la FFT discrète, $\Delta f$, ne doit pas dépasser 0,05 Hz (Rathje et. al, 2004).

2.2.2. *Paramètres basés sur les spectres de réponse*

(i) Dans la littérature on utilise aussi, pour l'évaluation de la période prédominante, les périodes qui correspondent à la valeur maximale des ordonnées des spectres de l'accélération, $T_{gSA}$ (Rathje et al., 1998), de la vitesse, $T_{gSV}$ (Miranda, 1993a) ou de l'énergie introduite dans le système, $T_{gEI}$, (Miranda, 1993b). Tous ces spectres sont calculés pour une fraction d'amortissement critique de 5%.

(j) La période caractéristique, $T_1^*$, est définie comme la période de transition entre le segment d'accélération constante et le segment de vitesse constante d'un spectre élastique. La période caractéristique est donnée par la relation

$$T_1^* = 2\pi \frac{(s_v)_{\max}}{(s_a)_{\max}} \quad [9]$$

où $(s_v)_{\max}$ et $(s_a)_{\max}$ sont les ordonnées maximales des spectres de la pseudo-vitesse et de la pseudo-accélération, calculés pour une fraction amortissement critique de 5%.

Une définition alternative de la période caractéristique a été propose par Lungu et al. en 1997 (Dubina et Lungu, 2003), en utilisant les valeurs effectives modifiées de la vitesse et l'accélération du sol, EPV et EPA :

$$T_C = 2\pi \, EPV/EPA \quad [10]$$

avec

$$EPV = \left( \overline{SV}_{0,4\,s} \right)_{\max} / 2,5 \quad [11]$$

$$EPA = \left( \overline{SA}_{0,4\,s} \right)_{\max} / 2,5 \quad [12]$$

et où $\left( \overline{SV}_{0,4\,s} \right)_{\max}$ et $\left( \overline{SA}_{0,4\,s} \right)_{\max}$ sont les valeurs maximales des spectres de vitesse et d'accélération sur lesquelles on a appliqué une moyenne glissante avec une fenêtre mobile de période de 0,4 s. Dans la présente étude, on a utilisé cette seconde définition de $T_C$.

2.2.3. *Paramètres basées sur les valeurs maximales du mouvement du sol*

(k) Une relation empirique pour l'évaluation de la période caractéristique du mouvement du sol, proposée par Heidebrecht en 1987, est utilisée par Fajfar et Fischinger dans (Fajfar et al., 1990) :

$$T_{4.3} = 4.3 \frac{PGV}{PGA} \qquad [13]$$

Dans la formule [13], PGV et PGA sont, respectivement, la vitesse et l'accélération maximale du sol. Cette période a été utilisée pour définir les limites du domaine des périodes moyennes, à savoir le domaine de période où le spectre lissé de la pseudo-vitesse prend ses valeurs maximales. Selon les auteurs cités, l'expression de la période $T_{4.3}$ a résulté de l'hypothèse que l'amplification spectrale dans la région contrôlée par l'accélération du spectre de type Newmark-Hall est 1,46 fois plus grande que l'amplification dans la région contrôlée par la vitesse.

### 2.3. Etudes de corrélation

On a étudié les corrélations entre les paramètres de type période présentés dans la section précédente, et aussi entre ces paramètres et les indicateurs de la largeur de bande de fréquence.

Dans le tableau 1, on présente les valeurs des coefficients de corrélation, $R^2$, calculées pour tous les composantes horizontales des enregistrements considérées. Les valeurs sont déterminées dans l'hypothèse que la relation entre deux paramètres est donnée par une fonctionnelle linéaire $y=a \cdot x+b$.

**Tableau 1.** *Coefficients de corrélation ($R^2$) entre les paramètres analysés. Composantes horizontales pour tous les événements sismiques considérés*

|  | $T_{ms}$ | $T_{1(DSP)}$ | $T_{mean}$ | $T_{cen}$ | $T_{gSA}$ | $T_{gSV}$ | $T_{gSEI}$ | $T_C$ | $T_{4.3}$ | $q$ | $\varepsilon$ |
|---|---|---|---|---|---|---|---|---|---|---|---|
| $T_{ms}$ | **1.000** | **0.531** | **0.763** | **0.570** | *0.441* | *0.285* | *0.417* | **0.831** | **0.759** | **0.515** | *0.445* |
| $T_{1(DSP)}$ | **0.531** | **1.000** | *0.384* | *0.261* | *0.227* | *0.189* | *0.358* | *0.468* | *0.378* | *0.324* | *0.161* |
| $T_{mean}$ | **0.763** | *0.384* | **1.000** | **0.921** | **0.543** | *0.127* | *0.204* | **0.512** | **0.559** | *0.272* | *0.481* |
| $T_{cen}$ | **0.570** | *0.261* | **0.921** | **1.000** | *0.443* | 0.071 | *0.126* | *0.341* | *0.410* | 0.098 | *0.420* |
| $T_{gSA}$ | *0.441* | *0.227* | **0.543** | *0.443* | **1.000** | 0.030 | 0.064 | *0.276* | *0.307* | *0.196* | *0.356* |
| $T_{gSV}$ | *0.285* | *0.189* | *0.127* | 0.071 | 0.030 | **1.000** | *0.417* | *0.366* | *0.266* | *0.253* | *0.127* |
| $T_{gSEI}$ | *0.417* | *0.358* | *0.204* | *0.126* | 0.064 | *0.417* | **1.000** | *0.426* | *0.318* | *0.258* | *0.111* |
| $T_C$ | **0.831** | *0.468* | **0.512** | *0.341* | *0.276* | *0.366* | *0.426* | **1.000** | **0.715** | *0.442* | *0.274* |
| $T_{4.3}$ | **0.759** | *0.378* | **0.559** | *0.410* | *0.307* | *0.266* | *0.318* | **0.715** | **1.000** | *0.379* | *0.310* |

Si on considère $R^2 > 0,5$ comme correspondant à une bonne corrélation (police en gras dans le tableau), $R^2 = 0,1...0,5$ comme une corrélation modérée (en italique) et $R^2 < 0,1$ comme une corrélation faible (en maigre), on peut observer que, pour les paramètres de période analysés, on trouve des bonnes corrélations en 28 % des cas, des corrélations modérées en 64 % des cas et des faibles corrélations en 8% des cas.

En ce qui concerne la corrélation entre les valeurs de période et les valeurs des indicateurs de la largeur de bande de fréquences, $q$ et $\varepsilon$, on constate que les corrélations sont plutôt modérées.

Le tableau 1 montre aussi des différences significatives concernant les degrés de corrélation entre les paramètres examinés.

En analysant la période moyenne quadratique, $T_{ms}$, un paramètre considéré comme l'un des plus prometteurs pour exprimer le contenu en fréquences des mouvements du sol (Rathje et al., 2004), on trouve qu'il a la meilleure corrélation ($R^2$=0,831) avec la période caractéristique basée sur les valeurs effectives modifiées du mouvement du sol, $T_C$.

Des bonnes corrélations ($R^2 \cong 0,76$) sont aussi observées entre $T_{ms}$ et $T_{mean}$, d'un part, et entre $T_{ms}$ et $T_{4.3}$, d'un autre part. On remarque, également, des bonnes corrélations, quoique caractérisées par des coefficients $R^2$ plus bas (0,570 et 0,531), entre $T_{ms}$ et la période centrale, $T_{cen}$, et entre $T_{ms}$ et $T_{1(DSP)}$. La période $T_{ms}$ est assez bien corrélée avec les indicateurs de la largeur de la bande de fréquences, $q$ est ε.

En ce qui concerne les autres paramètres du type période, on observe une très bonne corrélation entre la période moyenne, $T_{mean}$, et la période centrale, $T_{cen}$ ($R^2=0,921$), qui ont tous les deux des définitions basées sur les moments spectraux de la DSP. Des autres corrélations à remarquer sont celles entre $T_C$ et $T_{4.3}$ ($R^2=0,715$), $T_{4.3}$ et $T_{mean}$ ($R^2=0,559$), $T_{gSA}$ et $T_{mean}$ ($R^2=0,543$) et aussi entre $T_C$ et $T_{mean}$ ($R^2=0,512$).

La période correspondant au maximum du DSP, $T_{1(DSP)}$, est modérément corrélée avec les autres paramètres analysés, à l'exception de la période $T_{ms}$, avec laquelle elle a une bonne corrélation, comme mentionné ci-dessus.

A l'autre extrême, les plus faibles corrélations sont celles entre $T_{gSA}$ et $T_{gSV}$ ($R^2=0,030$) et entre $T_{gSA}$ et $T_{gSEI}$ ($R^2=0,064$).

Les corrélations ont été également examinées, séparément, pour chaque événement sismique considéré. Dans la figure 2, on présente des illustrations des résultats obtenus.

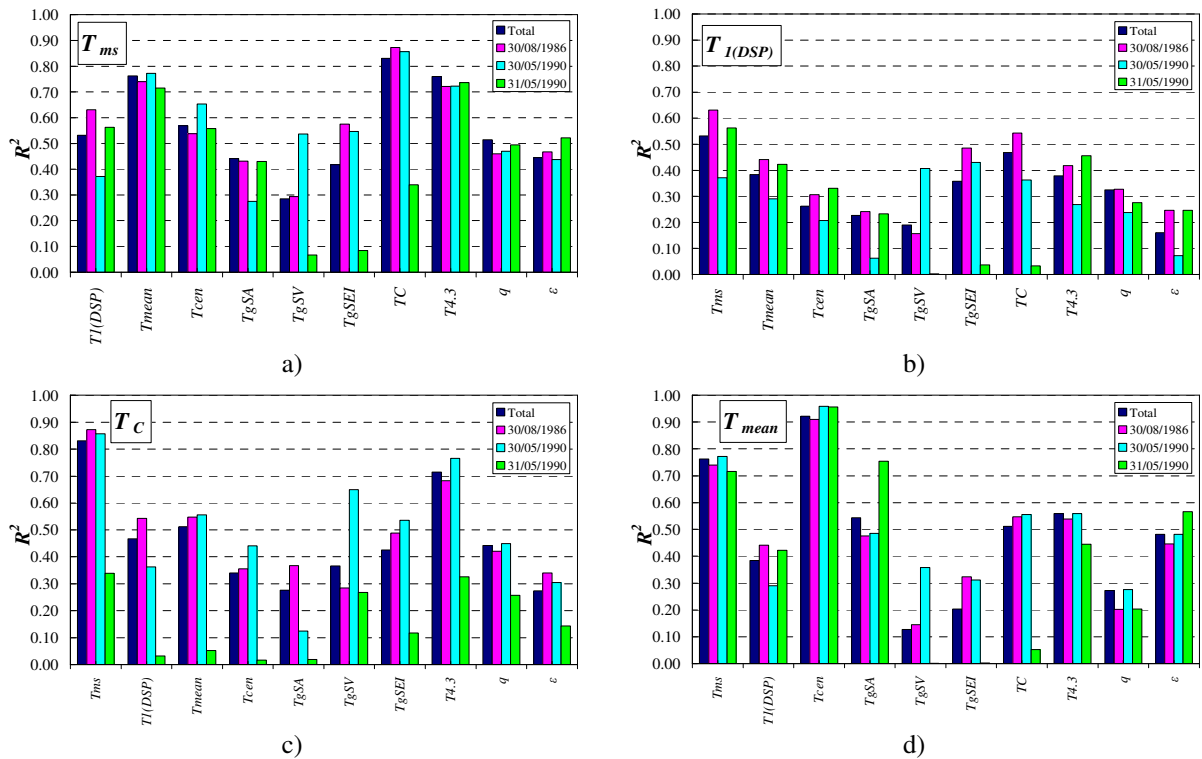

**Figure 2.** *Corrélations entre différents paramètres analysés, pour l'ensemble des enregistrements considérées et pour chaque événement sismique*

On remarque, dans la figure 2, pour une grande partie des corrélations examinées, une variation accentuée des coefficients de corrélation avec l'événement sismique considéré. Pourtant, on peut aussi observer un nombre de corrélations plus stables, comme $T_{ms} - T_C$, $T_{ms} - T_{mean}$, $T_{ms} - T_{4.3}$, $T_{ms} - q$, $T_{ms} - \varepsilon$, $T_{mean} - T_{cen}$, $T_{mean} - T_{4.3}$, $T_{mean} - \varepsilon$.

Si on détermine la corrélation entre les paramètres dans l'hypothèse que la relation entre eux est décrite par une fonctionnelle linéaire $y=a \cdot x$, on obtient les résultats présentés dans la figure 3. On a représenté les paramètres avec les corrélations les plus stables par rapport à l'événement sismique, identifiées ci-dessus.

On constate que, dans la figure 3, les lignes de tendance sont presque superposées, en suggérant des dépendances simples et stables entre les paramètres considérés. A remarquer que les paramètres $T_{ms}$, $T_C$, $T_{mean}$ et $T_{4.3}$ ont des définitions très différentes, ce que donne un plus de signifiance aux bonnes corrélations existant entre eux.

Une sélection supplémentaire parmi les 218 composantes horizontales examinées, basée sur leur largeur de bande de fréquences, a identifié 21 composantes à bande étroite, enregistrées lors des séismes des 4 mars 1977, 30 août 1986 et 30 mai 1990. Ce nombre a été considéré insuffisant pour obtenir des conclusions fiables par des études de corrélation.

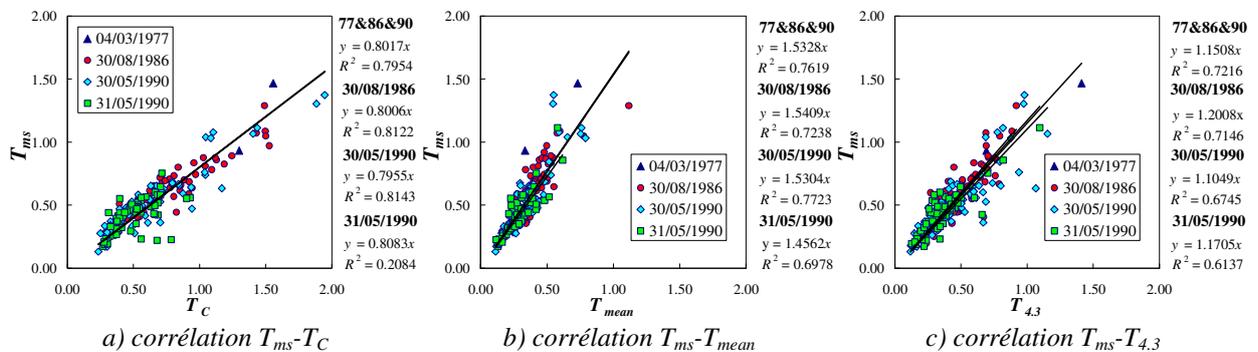

*a) corrélation $T_{ms}$-$T_C$*   *b) corrélation $T_{ms}$-$T_{mean}$*   *c) corrélation $T_{ms}$-$T_{4.3}$*

**Figure 3.** *Corrélations entre la période $T_{ms}$ et des autres paramètres analysés*

On a effectué, aussi, une étude préliminaire sur les composantes verticales des accélérogrammes, en constatant une corrélation améliorée entre $T_{I(DSP)}$ et les périodes $T_{gSV}$ et $T_{gSEI}$, par comparaison aux résultats obtenu pour les composantes horizontales.

## 2.4. *Distribution spatiale des paramètres calculés*

La distribution spatiale des différents paramètres considérés a été examinée, en cartographiant les valeurs calculées. Tenant compte des bons résultats obtenus pour $T_{ms}$ dans l'analyse de la section précédente, on présente, dans la figure 4, les cartes avec les valeurs de ce paramètre, pour les séismes de 1986 et 1990.

On observe des différences entre la distribution des valeurs de $T_{ms}$ lors des 3 séismes analysés, ainsi que la variation des valeurs de ce paramètre, pour la même station et pour les 3 séismes consécutifs.

Ces résultats seront analysés, dans les phases suivantes de la recherche, en corrélation avec les informations obtenues par l'étude individualisée des informations disponibles pour chaque station sismique.

### 3. Conclusions

On a étudié la capacité de neuf paramètres de type période, utilisées dans la littérature, à décrire le contenu en fréquences des mouvements sismiques. La recherche a utilisé plus de 300 accélérogrammes enregistrées lors des séismes forts, avec magnitude de moment supérieure à 6, qui se sont produits en Vrancea dans les 34 dernières années. De plus, on a utilisé aussi l'information fournie par les indicateurs de la largeur de bande de fréquences.

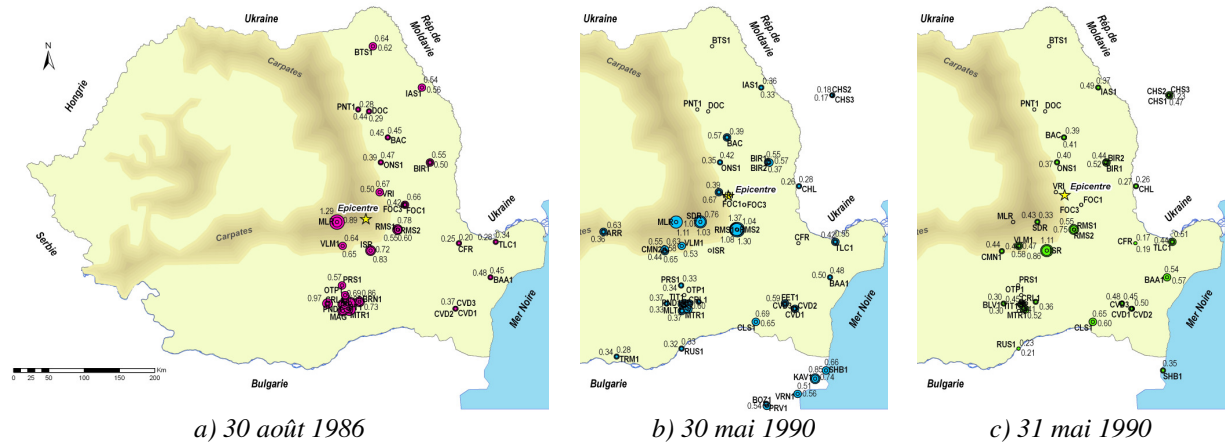

*a) 30 août 1986*   *b) 30 mai 1990*   *c) 31 mai 1990*

**Figure 4.** *Cartes des valeurs de $T_{ms}$, pour les séismes de 30 août 1986, 30 mai 1990 et 31 mai 1990*

Les corrélations entre les paramètres considérés ont été déterminées en considérant l'ensemble d'accélérogrammes, ainsi que les sous-ensembles fondés sur les critères : événement sismique, type de composante (horizontale ou verticale) et largeur de la bande de fréquences.

Un des paramètres les plus prometteurs semble être la période moyenne quadratique, proposée par Rathje et al. en 1998. Ce paramètre, défini comme une fonction des amplitudes Fourier et des fréquences, a démontré des corrélations bonnes et relativement stables avec les autres paramètres de type période examinés, ainsi qu'avec les indicateurs de la largeur de bande de fréquences. Autres paramètres d'intérêt sont la période caractéristique basée sur des définitions modifiées des valeurs maximales effectives du mouvement du sol, proposée par Lungu et al. en 1997, la période prédominante proposée par Heidebrecht en 1987 et, possiblement, la période moyenne et la période centrale, basées sur les moments spectraux de la densité spectrale de puissance.

On a étudié aussi la distribution spatiale des paramètres calculés, en réalisant des cartes pour chaque séisme analysé et en mettant en évidence la variation des valeurs d'un événement sismique á l'autre.

Des recherches supplémentaires sont nécessaires pour mieux déceler l'influence de divers facteurs (effet du site, chemin de propagation, mécanisme de foyer) sur le contenu en fréquences des mouvements sismiques analysées.

## 4. Bibliographie